\documentclass[11pt]{article}
\usepackage{moriond,epsfig}
\usepackage{amssymb}
\usepackage{mathrsfs}

\def\be{\begin{equation}}
\def\ee{\end{equation}}
\def\bea{\begin{eqnarray}}
\def\eea{\end{eqnarray}}

\begin{document}
\vspace*{4cm}
\title{DIRECTIONAL DETECTION OF DARK MATTER}

\author{ J. BILLARD, F. MAYET, J. F. MACIAS-PEREZ, D. SANTOS, C. GRIGNON, O. GUILLAUDIN }

\address{Laboratoire de Physique Subatomique et de Cosmologie, Universit\'e Joseph Fourier Grenoble 1,
  CNRS/IN2P3, Institut Polytechnique de Grenoble, 53 rue des Martyrs 28026 Grenoble, France}

\maketitle\abstracts{
Directional detection of galactic Dark Matter is a promising  search strategy for discriminating
genuine WIMP events from  background ones. However,  to take full advantage of this powerful detection method, one need to be able to 
  extract information from an observed recoil map to identify a WIMP signal. We present a comprehensive formalism, using a map-based 
  likelihood method 
  allowing to recover the main incoming direction of the signal, thus proving its galactic origin, and the corresponding significance.
   Constraints are then deduced in the ($\sigma_n, m_\chi$) plane. }

\section{Introduction}

\begin{figure}[t]
\begin{center}
\includegraphics[scale=0.2,angle=90]{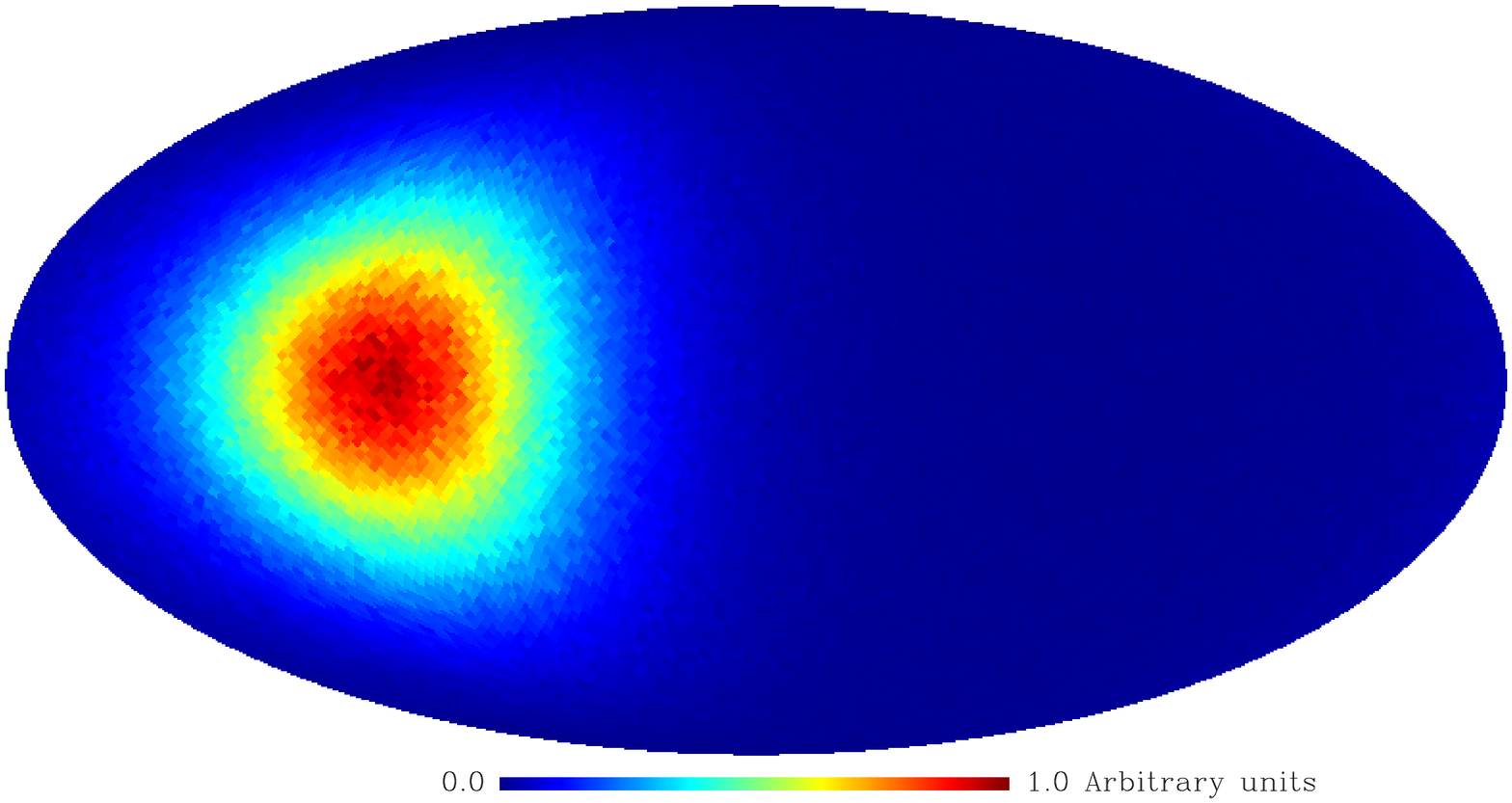}
\includegraphics[scale=0.2,angle=90]{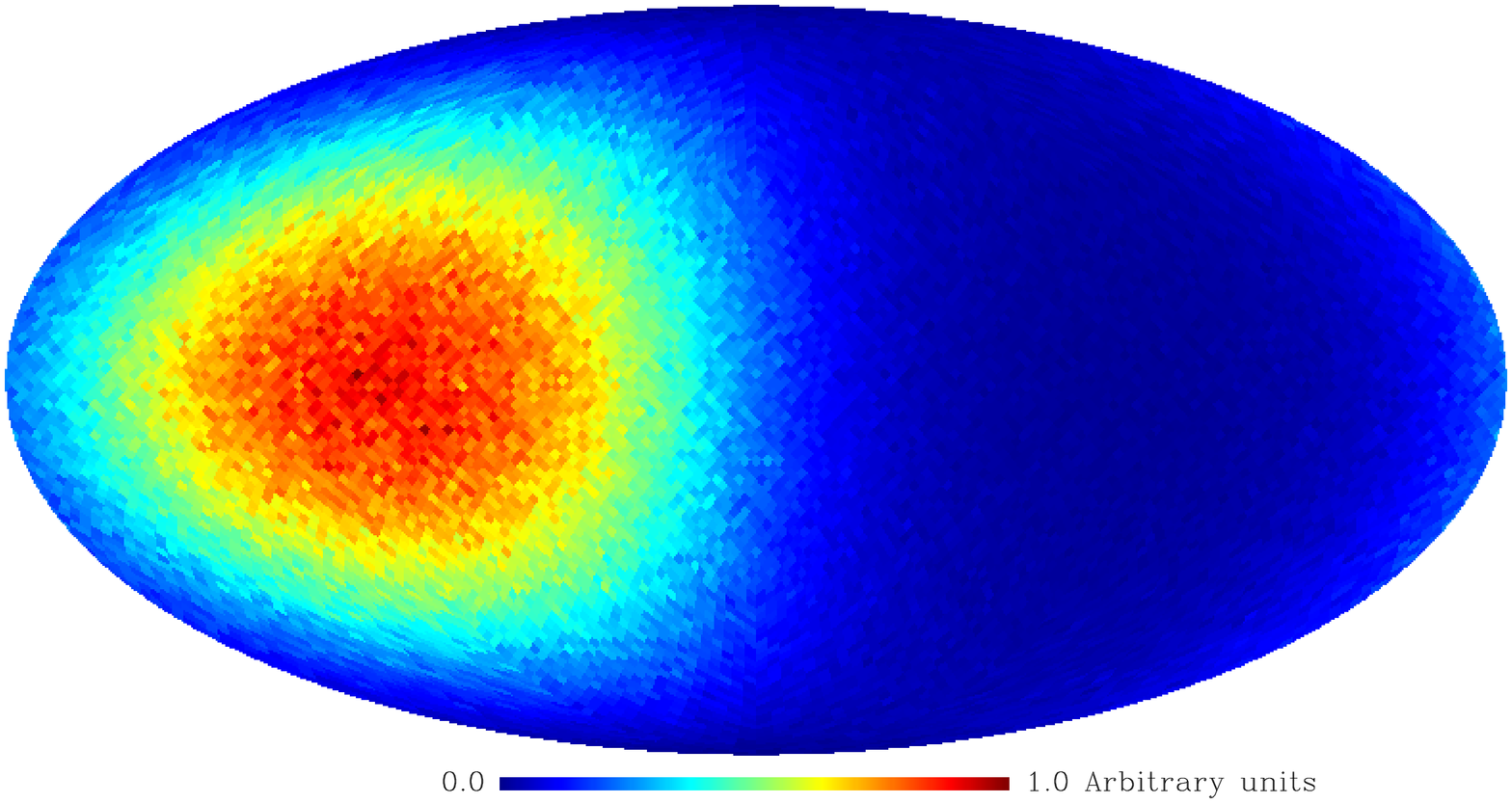}
\includegraphics[scale=0.2,angle=90]{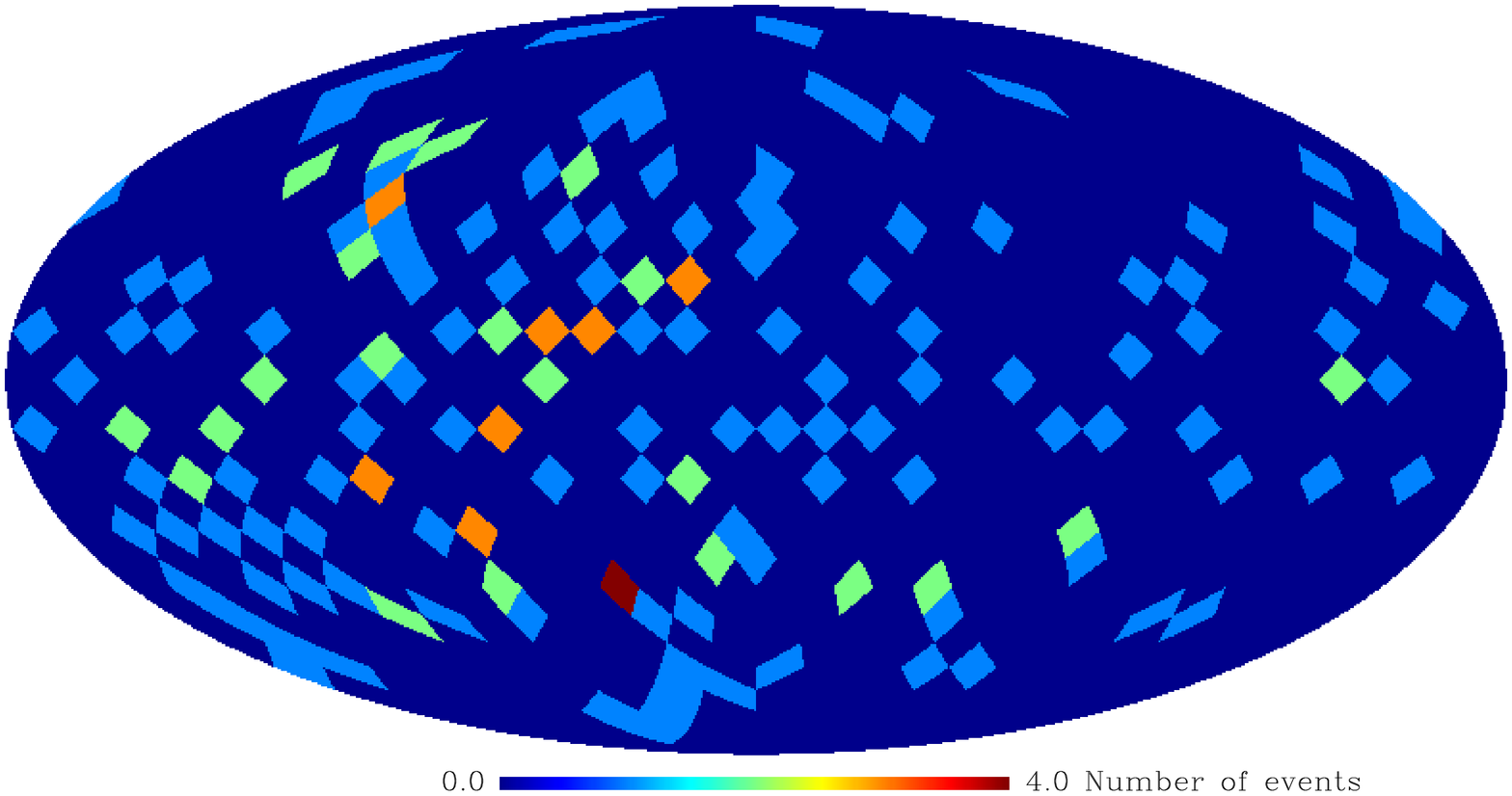}
\caption{From left to right : WIMP flux  in the case of an isothermal spherical halo,   WIMP-induced recoil distribution 
 and a typical simulated measurement :  100 WIMP-induced recoils and 100 background
events with a low angular resolution. Recoils maps are produced for a Fluorine target, a  100 GeV.c$^{-2}$ WIMP 
 and considering recoil energies in the range  5 keV $\leq E_R \leq$ 50 keV. Maps are Mollweide equal area projections.}  
\label{fig:DistribRecul}
\end{center}
\end{figure}

Taking advantage of the astrophysical framework, directional detection of Dark Matter is an interesting strategy in order to distinguish
 WIMP events from background ones.
Indeed, like most spiral galaxies, the Milky Way is supposed to be immersed in a halo of WIMPs which outweighs the luminous component by at 
least an order of magnitude. As the Solar System rotates around the galactic center through this Dark Matter halo, WIMPs should mainly come
 from the direction to which points the
Sun velocity vector and which happens to be roughly in the direction of the Cygnus constellation.
Then, a directional WIMP flux is expected to enter any terrestrial detectors (see fig.\ref{fig:DistribRecul} left) infering a directional
 WIMP
 induced recoil distribution which should be pointing toward the Cygnus Constellation, {\it i.e.} in the
  ($\ell_\odot = 90^\circ,  b_\odot =  0^\circ$) direction (see fig.\ref{fig:DistribRecul} middle).
  The latter corresponds to the expected WIMP signal probed by directional detectors and as it is shown on the fig.\ref{fig:DistribRecul}
  (middle), a strong anisotropy is expected \cite{spergel} while the background should be isotropic.\\
 
 Several project of directional detectors are being developed \cite{white} and in this paper,
   we present a map-based likelihood analysis \cite{billard} in order to extract from an 
observed recoil map the main incoming direction of the events and its significance. This way, the galactic 
origin of the signal, thus the identification of a genuine WIMP signal, can be proved by showing its correlation with the direction of 
the solar motion. This blind analysis is intended to be applied to directional data of any detector and as an example we will apply this
 method to a realistic simulated data.

\section{Map-based likelihood analysis}

\subsection{A realistic simulated measurement}
Right panel of figure \ref{fig:DistribRecul} 
presents a typical recoil distribution observed by a directional detector : $100$ WIMP-induced events and 
$100$ background events generated isotropically.  These   events  are meant to be after data rejection based e.g. on track length and energy
 selection\cite{santos}. 
For an elastic axial cross-section on nucleon $\rm \sigma_{n} = 1.5 \times 10^{-3} \ pb$ and a $\rm 100 \ GeV.c^{-2}$ WIMP mass, this
 corresponds to 
an exposure of $\rm \sim 7\times 10^3  \ kg.day$ in  $\rm ^{3}He$ and $\rm \sim 1.6 \times 10^3 \ kg.day$  in CF$_4$, 
on their equivalent energy ranges as discussed in \cite{billard}.
  Low resolution maps are used in this case ($N_{\rm pixels} = 768$) which is sufficient  for the 
low  angular resolution, $\sim 15^\circ$ (FWHM), expected for this type of detector and   justified  for instance by the  
straggling of the recoiling nucleus\cite{santos}. 
3D read-out and sense recognition are considered.\\

\subsection{Likelihood definition}
At first sight, it seems difficult to conclude from the recoil map of fig.~\ref{fig:DistribRecul} (right) that it does contain 
a fraction of WIMP events pointing towards the direction of the solar motion. 
A likelihood analysis is developed in order to retrieve from a recoil map : the main direction of the incoming events in 
galactic coordinates ($\ell, b$) and the number of WIMP events contained in the map. The likelihood value is estimated using a binned map 
of the overall sky with  Poissonian statistics,  as follows :
 \begin{equation}
 \mathscr{L}(m_\chi,\lambda, \ell,b) = \prod_{i=1}^{N_{\rm pixels}} P(  \lambda S_i(m_\chi ;\ell,b) + (1-\lambda) B_i |M_i)
 \end{equation}
where $B$ is the  background spatial distribution 
taken as isotropic, $S$ is the WIMP-induced recoil distribution and $M$ is a typical measurement. 
This is a four parameter likelihood analysis with $m_\chi$, 
 $\lambda = S/(B+S)$ the  WIMP fraction (related to the  background 
rejection power of the detector) and the angles ($\ell$, $b$) corresponding to the coordinates of the maximum of the 
expected WIMP events angular distribution.
Hence, $S(m_\chi;\ell,b)$ corresponds to a rotation of the $S(m_\chi)$ distribution 
by the angles ($\ell' = \ell - \ell_\odot$, $b' = b - b_\odot$).

A scan of the four parameters with flat priors, allows to evaluate the likelihood between the measurement 
(fig.~\ref{fig:DistribRecul} right) and the theoretical distribution made of a superposition of 
an isotropic background and a pure WIMP signal (fig. \ref{fig:DistribRecul} middle). By scanning on $\ell$ and $b$ values, we ensure 
that there is no prior on the direction of the center of the WIMP-induced recoil distribution. In order to respect the spherical topology,
 a careful rotation of the $S$ distribution
 on the whole sphere must be done as follows.
 Given $\overrightarrow{V}_i$ the vector pointing on a 
bin $S_i$, the following rotation is considered : 
$$\overrightarrow{V}^\prime_i = R_{\vec{u}}(b')R_{\hat{z}}(\ell')\overrightarrow{V}_i$$
 with $\vec{u} = R_{\hat{z}}(\ell') \ \hat{x} = u_x \ \hat{x} + u_y \ \hat{y}$ and $R_{\vec{u}}(b')$  is the  Rodrigues 
 rotation matrix around an arbitrary 
vector  $\vec{u}$.

The events contained in the observed recoil map can be either  a WIMP-induced recoil or a background event. A realistic recoil map from
upcoming directional detectors can not be background free. With this method, both components (background and signal) are taken into account
 and  
no assumption on the origin of each event is needed. Indeed, the observed map  is considered as a superposition of  
the background and WIMP signal distribution, and the likelihood method allows to recover $\lambda$, the signal to noise ratio. 
The advantage
is twofold :
\begin{itemize} 
\item Firstly, background-induced bias is avoided. This would not be the case with a method trying to 
evaluate a  likelihood on a map containing a fairly large number of background events considering only a pure WIMP reference distribution.  
\item Secondly, the value of $\lambda$ allows to access the number of genuine WIMP events
and consequently the scaterring cross-section as presented in sec. \ref{sec:cross}.
\end{itemize}

It is worth noticing that the likelihood is performed on the whole angular distribution in order to maximize the information contained 
in an observed recoil map, to be sensitive to the Dark Matter halo shape and to enhance the constraint on the four parameters. 

\subsection{Results from a realistic recoil map}

 \begin{figure}[t]
\begin{center}
\includegraphics[scale=0.16,angle=270]{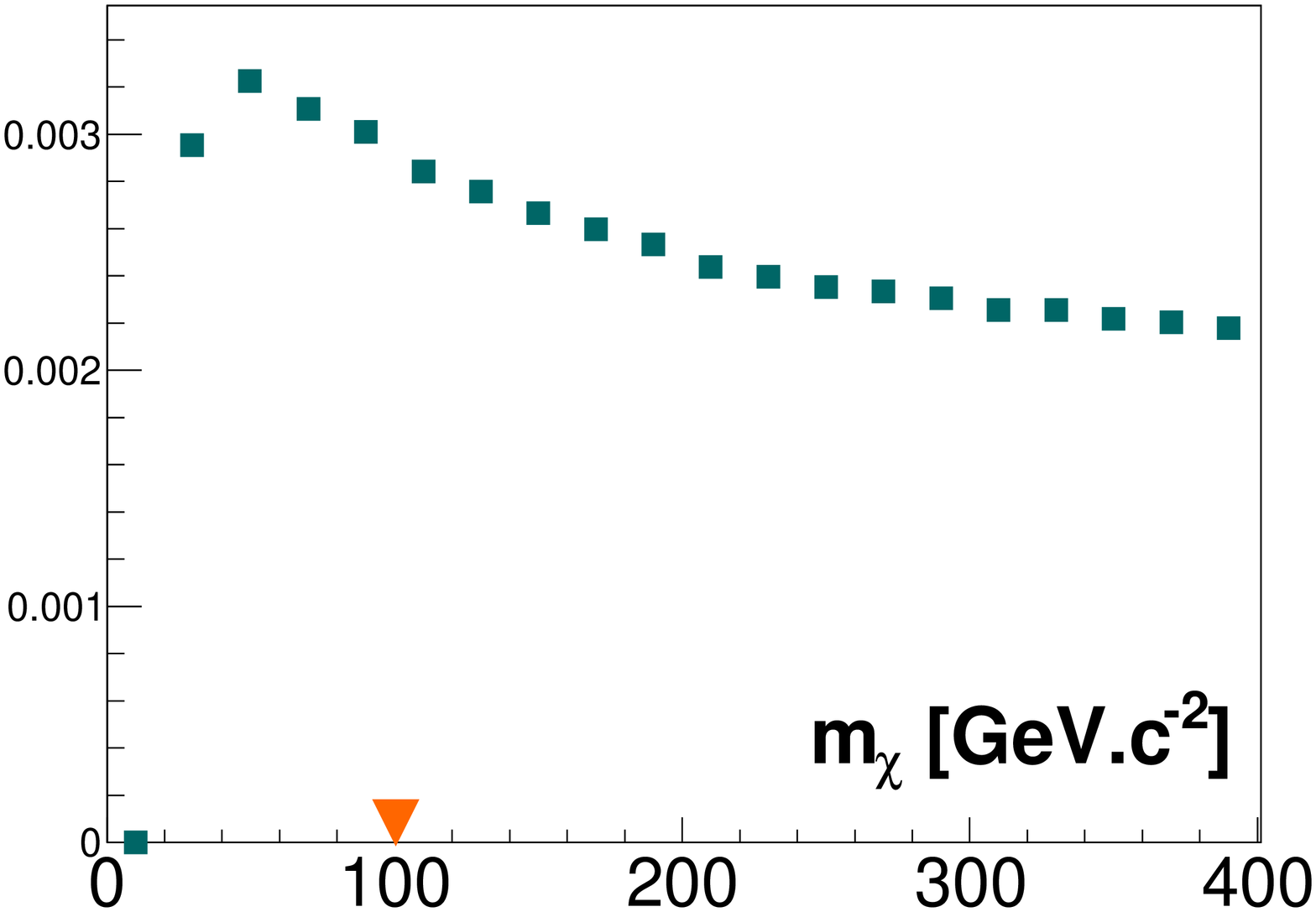}
\hspace*{1mm}
\includegraphics[scale=0.16,angle=270]{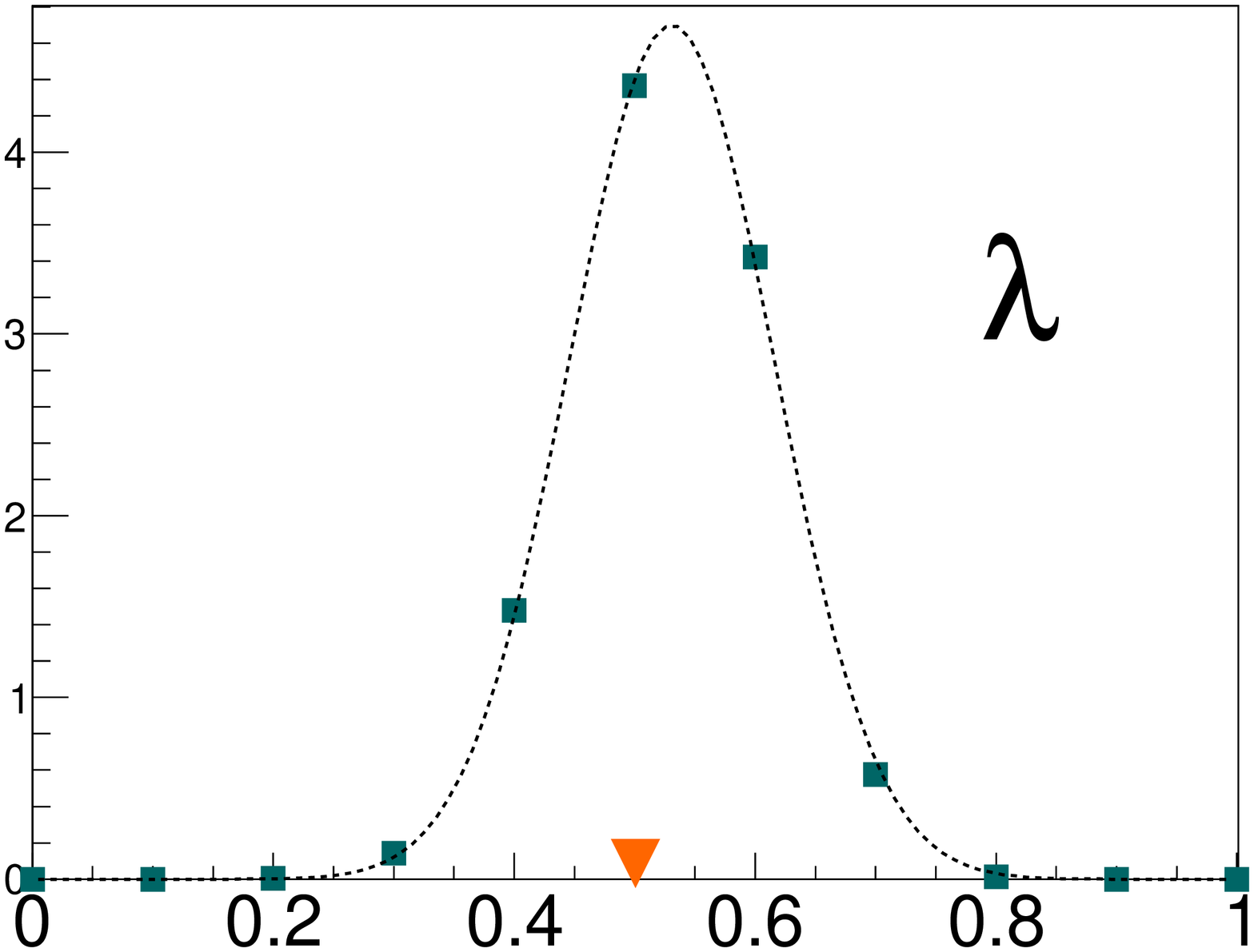}
\hspace*{1mm}
\includegraphics[scale=0.16,angle=270]{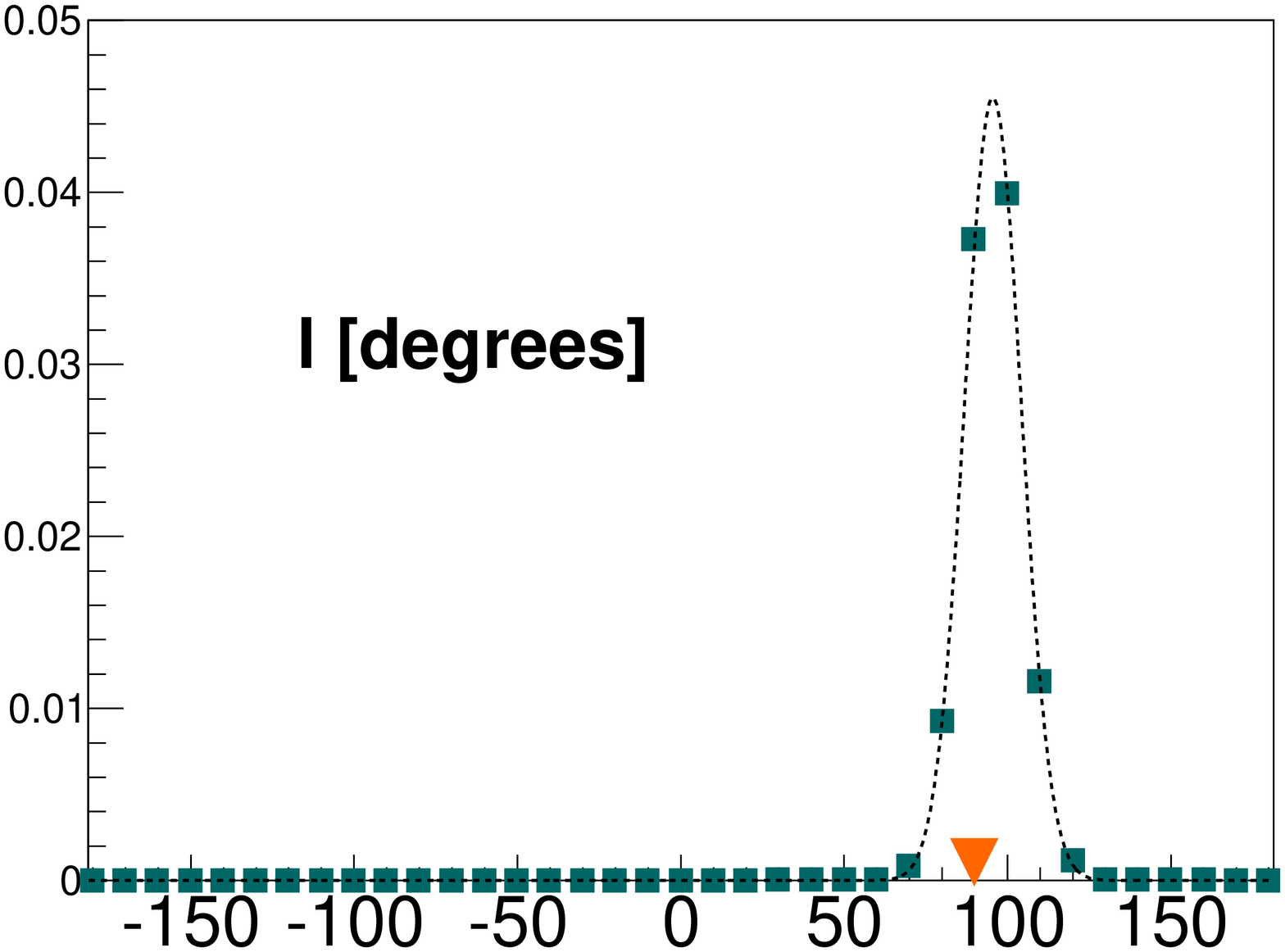}
\hspace*{1mm}
\includegraphics[scale=0.16,angle=270]{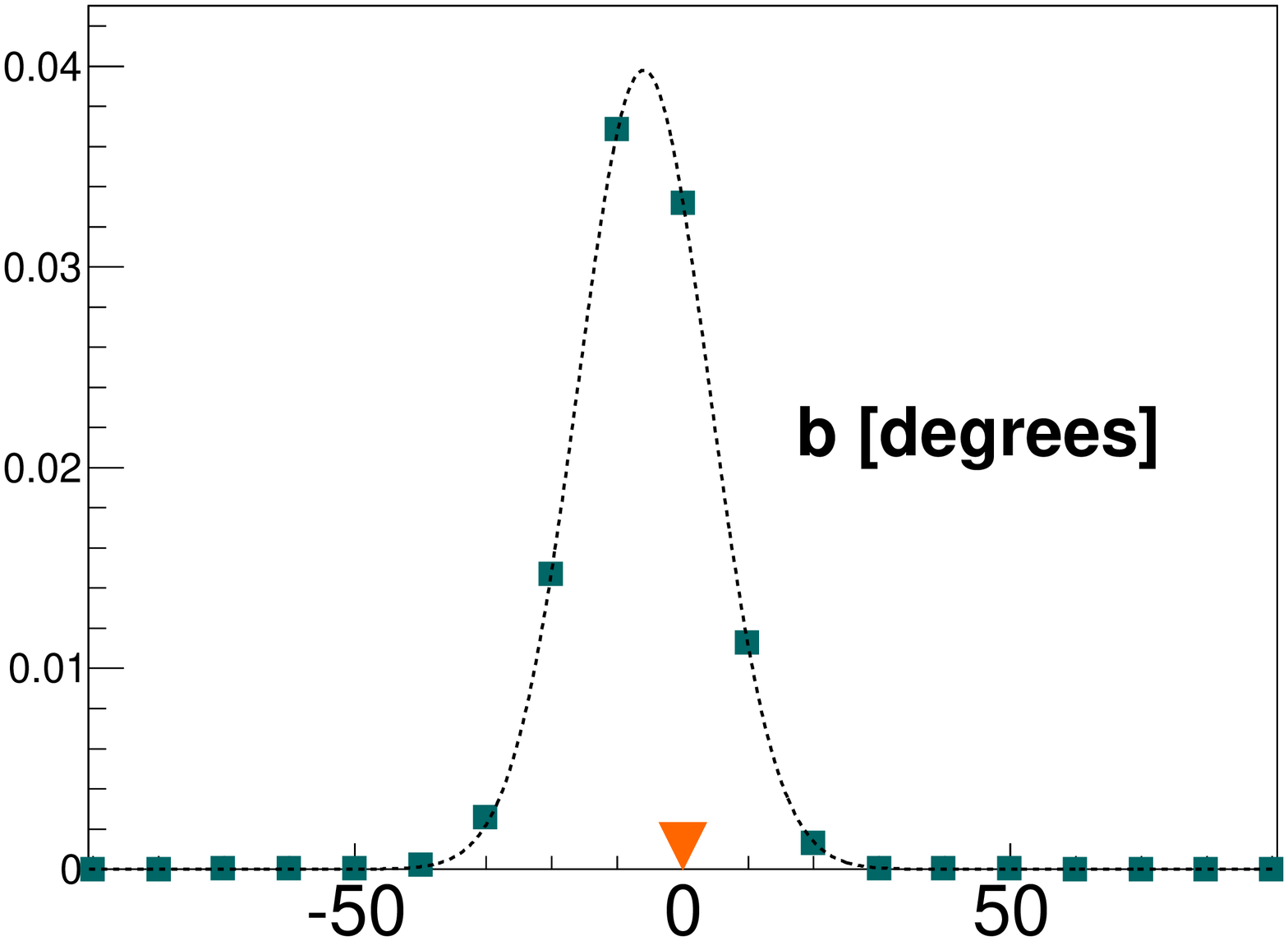}
\caption{From left to right : Marginalised probability density functions of the following parameters: $m_{\chi}$, $\lambda$, $\ell$ and $b$ after the likelihood analysis of the
simulated recoil map of fig.~\ref{fig:DistribRecul} right.}  
\label{fig:DistribParametres}
\end{center}
\end{figure}

The four parameter likelihood analysis has been computed on 
the simulated map (fig. \ref{fig:DistribRecul} right) and the marginalised distributions plots  of the four 
parameters $m_\chi , \lambda, \ell, b$ are presented in figure (\ref{fig:DistribParametres}).
 The conclusion of the analysis is threefold:
 \begin{itemize}
\item Firstly, as the $\ell$ and $b$ parameters are well constrained, the first result of this  map-based likelihood method is that  
the main recoil direction is recovered and it is pointing towards 
($ \ell = 95^{\circ} \pm 10^{\circ}, b = -6^{\circ} \pm 10^{\circ}$) at 
$68 \  \%$ CL, corresponding to a non-ambiguous detection of particles from the galactic halo. This is indeed the discovery  
proof of this detection strategy.
\item Secondly, the method allows to constrain the $\lambda$ parameters, the WIMP fraction, and then to derive the number of WIMP 
events contained in the observed recoil map. Indeed, we can estimate the number of WIMP events as $N_{\rm wimp} = \lambda \times N_{tot}$
 where $N_{tot} = S + B$ follows a Poissonian statistic, and $\Delta N_{\rm wimp}$ is given by 
$\Delta N_{\rm wimp} \approx \Delta \lambda \times N_{tot}$. Hence for this simulated recoil map, the reconstructed number of WIMP
 events is 
$N_{\rm wimp}=106 \pm 17 \ (68 \% {\rm CL})$.
\item Thirdly, we can notice that the WIMP mass is not recovered, there is only a lower limit: $m_{\chi} > 10$ GeV.c$^{-2}$.
 In fact, $m_{\chi}$ is set as a free parameter in order to show that the analysis is particle physics model independent.

\end{itemize}

As a conclusion of this analysis, the combination of the two main previous results given by the analysis, which are that the recoil
 map contains a signal pointing toward the Cygnus constellation within 10$^\circ$ with
  $N_{\rm wimp}=106 \pm 17 \ (68 \% {\rm CL})$, leads to a high
significance detection of galactic Dark Matter.

\subsection{Constraining the elastic scattering cross-section}
\label{sec:cross}
 \begin{figure}[t]
\begin{center}
\includegraphics[scale=0.35,angle=270]{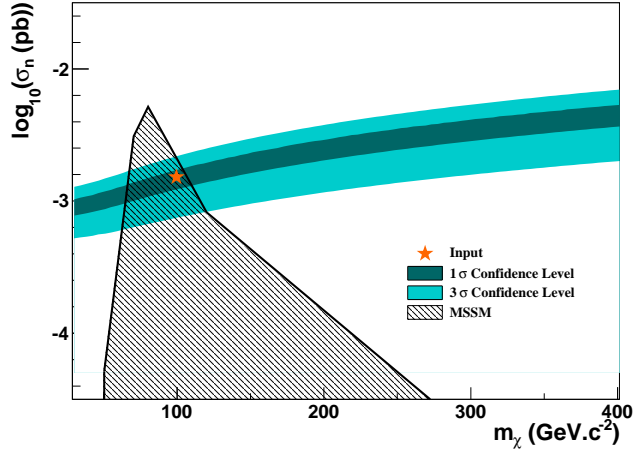}
\caption{Allowed regions obtained with the example map shown on figure \ref{fig:DistribRecul}. Results are presented in the plane
 : WIMP cross-section on nucleon ($\rm \sigma_n$) as a function of WIMP mass (GeV.c$^{-2}$), with $1\sigma$ and $3\sigma$
  CL contours. MSSM refers to generic SUSY models and the input value for the simulation is shown with a star.}
\label{fig:discovery}
\end{center}
\end{figure} 
A constraint in the $(\sigma_n, m_\chi)$ plane is then  deduced from  
the marginalised $\mathscr{L}(\lambda)$ distribution evaluated for each WIMP mass above $10$ GeV.c$^{-2}$. 
Using  the  standard expression of the event rate with a form factor $F^2(E_R)$ taken equal to one and the local halo density
 $\rho_0 = 0.3$ GeV.c$^{-2}$.cm$^{-3}$,
the $1\sigma$ and $3\sigma$  CL contours are calculated. Figure \ref{fig:discovery} presents the discovery region deduced from the
 analysis of the simulated recoil map. It should
be highlighted that these contours represent the allowed regions, as directional detection aims at identifying WIMP signal rather than
 rejecting the background.  
For reference, generic SUSY models \cite{ellis} 
are also shown as well as the input value for the simulation (orange star). Such a  result could be obtained, with a
 background rate of 
$\sim 0.07$ kg$^{-1}$day$^{-1}$ and a  10 kg $\rm CF_4$ detector 
during  $\sim 5$ months, noticing that the detector should allow 3D 
track reconstruction, with sense recognition down to 5 keV. 
  
\section{Conclusion}
\label{sec:conclusion}
We have presented a statistical analysis tool to 
extract information from  a  data sample of a directional detector in order to identify a galactic WIMP signal. 
As a proof of principle, it has been tested within the framework of an isothermal spherical halo model.
We have shown the feasibility to extract from an observed map the main incoming direction of the signal and its significance,
 thus  proving  its galactic origin. 
Systematical studies have been done\cite{billard} in order to show that this analysis tool gives satisfactory results on a large range of
exposure and background contamination.

\end{document}